\documentclass[sigconf]{acmart}

\usepackage{booktabs}

\usepackage[normalem]{ulem}  
\usepackage{bbold}
\usepackage{enumerate}
\usepackage{tabularx}
\usepackage{amsmath}
\usepackage{amssymb}
\usepackage[ruled,vlined,linesnumbered]{algorithm2e}
\usepackage{hyperref}
\usepackage{cleveref}
\usepackage{subfigure}
\usepackage{enumitem}

\DeclareMathOperator*{\argmax}{arg\,max}

\setcopyright{acmcopyright}

\acmDOI{10.1145/3234944.3234964}

\acmISBN{978-1-4503-5656-5/18/09}

\acmConference[ICTIR'18]{2018 ACM SIGIR International Conference on the Theory of Information Retrieval}{September 14--17, 2018}{Tianjin, China}
\acmYear{2018}
\copyrightyear{2018}

\acmPrice{15.00}

\begin{document}

\title[{Generating Synthetic Data for Neural Keyword-to-Question Models}]{Generating Synthetic Data for Neural Keyword-to-Question Models}

\author{Heng Ding}
\affiliation{%
  \institution{Wuhan University}
  \city{Wuhan}
  \country{China}
}
\email{hengding@whu.edu.cn}

\author{Krisztian Balog}
\affiliation{%
  \institution{University of Stavanger}
  \city{Stavanger}
  \country{Norway}
}
\email{krisztian.balog@uis.no}

\begin{abstract}
Search typically relies on keyword queries, but these are often semantically ambiguous.  We propose to overcome this by offering users natural language questions, based on their keyword queries, to disambiguate their intent.  This keyword-to-question task may be addressed using neural machine translation techniques.  Neural translation models, however, require massive amounts of training data (keyword-question pairs), which is unavailable for this task.  The main idea of this paper is to generate large amounts of synthetic training data from a small seed set of hand-labeled keyword-question pairs.  Since natural language questions are available in large quantities, we develop models to automatically generate the corresponding keyword queries.  Further, we introduce various filtering mechanisms to ensure that synthetic training data is of high quality.  We demonstrate the feasibility of our approach using both automatic and manual evaluation. This is an extended version of the article published with the same title in the Proceedings of ICTIR'18.

\end{abstract}



\maketitle

\section{Introduction}
\label{sec:intro}

\begin{figure}[t]
\centering
\vspace*{\baselineskip}
\includegraphics[width=8cm]{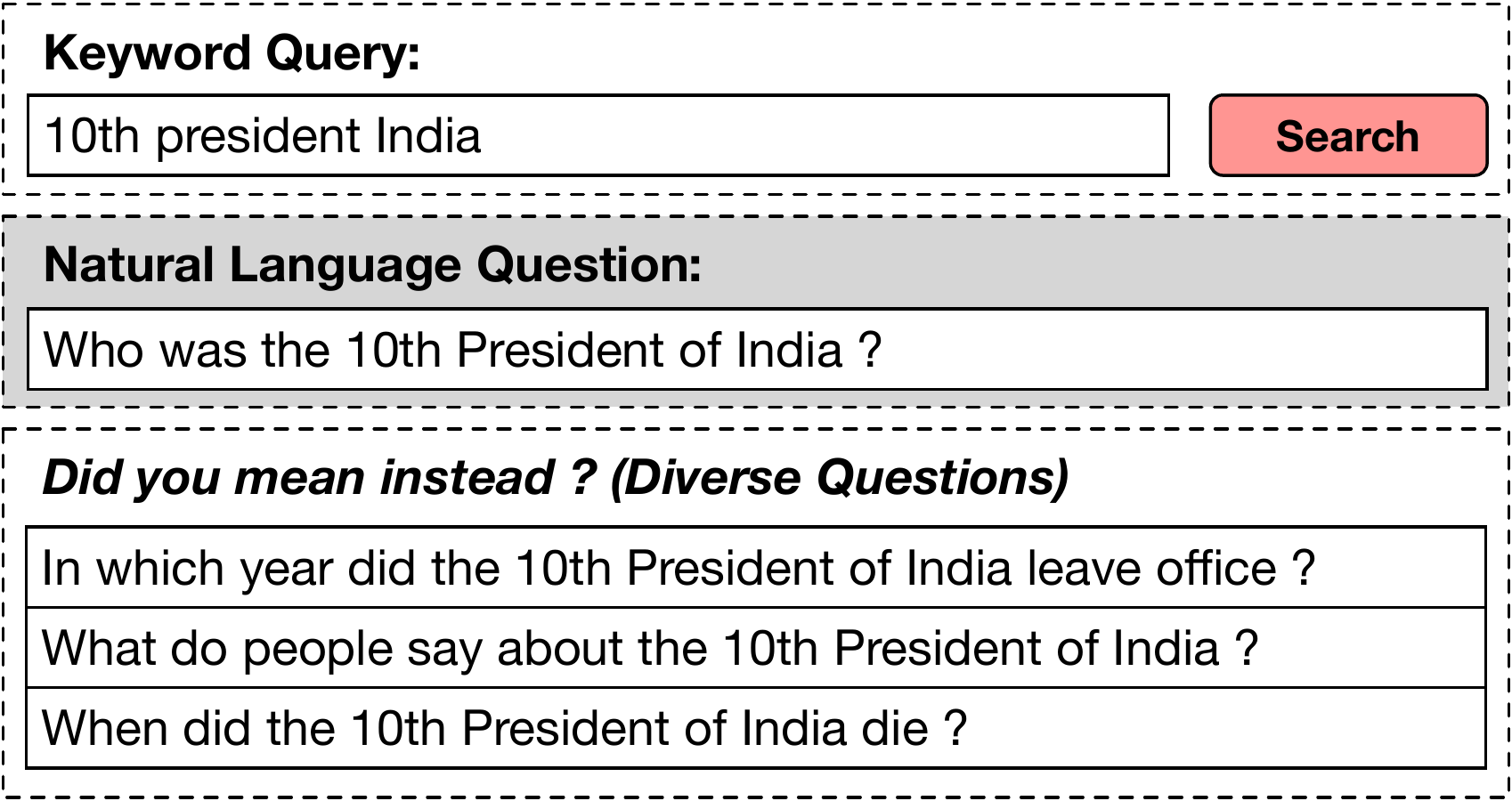}
\caption{Translating a keyword query to natural language question(s). 
Our focus is on the shaded area: generating the most common question for a keyword query.  The bottom part, generating diverse questions, is left for future work.}
\label{fig:ex1}
\vspace*{-0.5\baselineskip}
\end{figure}

Most search queries are motivated by some underlying question~\citep{Kotov:2010:TNQ}.  Today's users are accustomed to expressing the questions they have in mind using keyword queries~\cite{Zhao:2011:AGQ}.  Keyword queries, however, can be notoriously ambiguous and may be interpreted in multiple ways.   
For example, given the keyword query \emph{``10th president India,"} the question perhaps most users would want to ask is \emph{``Who was the 10th President of India?"}.  
Nevertheless, some users may be interested in a particular aspect of the query topic, like \emph{``In which year did the 10th President of India leave office?''} or \emph{``What do people say about the 10th President of India?''}.
By determining the underlying question, we can obtain a more accurate representation of the user's information need.  This, in turn, can lead to improved retrieval performance and a better overall search experience.
We envisage a search interface that allows users to refine their queries with automatically generated natural language questions; see Fig.~\ref{fig:ex1}.
We note that similar functionality is already offered, for certain queries, in major Web search engines (see Fig.~\ref{fig:googlex}).  Those services, however, are limited to suggesting existing questions to which answers are known to exist. 
Importantly, we are not aiming to retrieve existing questions from community question-answering archives~\citep{Xue:2008:RMQ,Gao:2013:MQQ}. 
Our goal is to automatically generate a natural language question that most likely represents the user's underlying information need.  
This is seen as a feedback mechanism that can more naturally engage users into explicitly clarifying their information needs.
How those natural language questions are actually utilized in a retrieval system (e.g., via query expansion~\citep{Kotov:2010:TNQ}) is beyond the scope of this study.

In this paper, we address the \emph{keyword-to-question} (K2Q) task: generating a natural language question from a keyword query.  
K2Q has generated considerable attention recently, see, e.g.,~\cite{Kotov:2010:TNQ,Dror:2013:QQO,Zhao:2011:AGQ,Zheng:2011:K2Q}.  Most existing works employ a template-based approach, where common question patterns are extracted from existing keyword-question pairs.  These template-based methods are inherently limited in their ability to generalize to previously unseen queries. 
Instead, we propose to address the K2Q task using state-of-the-art neural machine translation (sequence-to-sequence) approaches.  One challenge we face is that training such neural models requires massive amounts of training data (i.e., hand-labeled keyword-question pairs).
While such training data could be mined from query and click logs, there are two main issues.   First, such click data is not always available (e.g., in a cold start scenario).  Second, it is limited to keyword-question pairs that have received sufficiently many clicks; long-tail queries or newly posted questions will not have that.
The above considerations give rise to the main research objective of the present work: \emph{How can we generate synthetic data for training a neural machine translation approach for the K2Q task?}

The idea of generating synthetic data for training deep neural network has already been successfully applied for some computer vision tasks~\citep{Handa:2015:SUR, Zhang:2015:SUR, Gan:2015:LDS, Ros:2016:SDL}.  
In information retrieval, prior work has studied the creation of pseudo test collections, i.e.,  automatically generating query-document pairs, for training and evaluating retrieval algorithms~\citep{Azzopardi:2006:ACK, Azzopardi:2007:BSQ, Berendsen:2013:PTC}.
Inspired by those studies, we propose an approach that automatically generates large amounts of simulated keyword-question pairs from a small set of hand-labeled keyword question pairs, and then learns a neural keyword-to-question model with such synthetic training data.  The main technical contributions of this work are the following:

\begin{enumerate}[leftmargin=*]
  \item We present a novel approach for generating synthetic training data from a seed set of hand-labeled keyword-question pairs, and subsequently use this data for learning neural machine translation models to solve the K2Q task (Sect.~\ref{sec:overview}).
  \item We introduce several generative models for producing synthetic keyword queries from natural language questions (Sect.~\ref{sec:generation:kqgm}). 
  \item We develop two filtering mechanisms, which are essential for ensuring that the synthetic training data we feed into the neural network is of high-quality (Sect.~\ref{sec:generation:filter}).
  \item We evaluate our synthetic data generation approach on the end-to-end K2Q task using both automatic and manual evaluation (Sect.~\ref{sec:eval}).
\end{enumerate}

\section{Overview}
\label{sec:overview}

\begin{figure}[t]
\centering
\includegraphics[width=8cm]{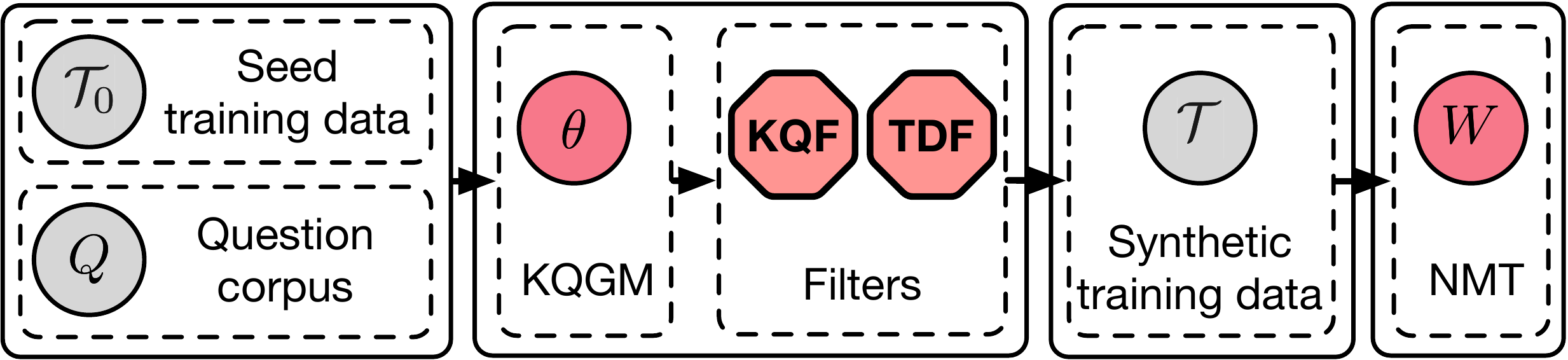}
\caption{The overview of our approach.  A small set of hand-labeled training data ($\mathcal{T}_0$) and a large scale question corpus ($\mathcal{Q}$) are input to the keyword query generation model (KQGM, parameterized by $\theta$). The output is passed through a keyword query filter (KQF) and a training data filter (TDF), resulting in a synthetic training data set ($\mathcal{T}$).  The synthetic training data is used for learning the parameters ($W$) of a neural machine translation model (NMT).}
\label{fig:fm}
\vspace*{-0.5\baselineskip}
\end{figure}

The overall goal in this paper is to tackle the keyword-to-question (K2Q) problem using neural networks.  I.e., the task is to translate a keyword query (referred to as \emph{keyword} for short) to a natural language question (\emph{question} for short).  To be able to use neural networks for this task, massive amounts of training data are needed.  The main idea of our paper is to use a small seed set of hand-labeled training data to generate large amounts of synthetic training data.  
Specifically, the \emph{seed training data}, $\mathcal{T}_0$, consists of keyword-question pairs, $(k,q) \in \mathcal{T}_0$.
This, along with a large \emph{question corpus}, $\mathcal{Q}$, is utilized to generate \emph{synthetic training data}, $\mathcal{T}$, which also consists of keyword-question pairs, $(k,q) \in \mathcal{T}$.
The neural machine translation models will then be trained using $\mathcal{T}$.
The overview of our framework is shown in Fig.~\ref{fig:fm}.  
It entails three main steps, which we shall detail below.

First, we train a keyword query generation model (KQGM), $\theta$, using keyword-question pairs from the seed training data.  We aim to simulate real users' querying behavior: given a natural language question, generate a keyword query that a user would likely issue when seeking an answer to that question.
We explore various generative models; these have only a few free parameters, which can be easily learned from the seed training data $\mathcal{T}_0$.  

Second, we utilize a large question corpus $\mathcal{Q}$, collected from community question answering forums, and employ the keyword query generation model $\theta$ to generate (a large set of) simulated keyword-question pairs.  These will constitute our synthetic training data $\mathcal{T}$.  
However, since not all the automatically generated keyword-question pairs are of high quality, we employ a keyword query filter (KQF) and a training data filter (TDF).  These filters are pivotal elements in our approach; we shall detail them in Sect.~\ref{sec:generation:filter}.

Finally, we train a neural machine translation (NMT) model for the K2Q task by feeding it with the synthetic training data $\mathcal{T}$.  We consider three neural networks: basic encoder-decoder NMT~\citep{Sutskever:2014:SSL}, NMT with attention mechanism~\citep{Bahdanau:2014:NMT}, and NMT with copying mechanism~\citep{Gu:2016:ICM}.  We shall detail these networks in Sect.~\ref{sec:nmt}.

\section{Synthetic Data Generation}
\label{sec:generation}

This section details the our synthetic training data generation method, which is the most important contribution of this paper.
The process takes as input (i) a small seed training data set, consisting of hand-labeled keyword-query pairs, and (ii) a large set of natural language questions.  The output is a large set of automatically generated keyword-question pairs, with high enough quality to train robust neural models.  Our approach consists of two main components: a keyword query generation model (Sect.~\ref{sec:generation:kqgm}) and filtering mechanisms (Sect.~\ref{sec:generation:filter}).

\subsection{Keyword Query Generation Model}
\label{sec:generation:kqgm}

Prior work has seen successful attempts at generating synthetic queries for web and microblog known-item search, both for evaluation and for training purposes~\citep{Azzopardi:2006:ACK, Azzopardi:2007:BSQ, Berendsen:2013:PTC}.
The overall idea is to construct a generative model that can produce a query, similar to a real query that a user would issue, for finding a particular item. 
We take the algorithm proposed by \citet{Azzopardi:2007:BSQ} as our starting point (\S\ref{sec:generation:kqgm:bl}) and extend it at several points to fit our problem setting: 
(i) we impose a number of restrictions as well as introduce new elements to the generative process (\S\ref{sec:generation:kqgm:modify}),
(ii) we propose a paraphrase-based variation that considers multiple ways of formulating the same question (\S\ref{sec:generation:kqgm:pqm}), and
(iii) we add phrase support, so as not to break up meaningful word sequences (\S\ref{sec:generation:kqgm:phrase}).

\subsubsection{Baseline}
\label{sec:generation:kqgm:bl}

In known-item search it is assumed that the user wants to find a particular item (document, question, tweet, etc.) that she has seen before in the corpus.  Therefore, the user constructs a keyword query by recalling terms that would help her identify this item.  In automatic query construction this user behavior is simulated using generative models.

Formally, let us assume that the user seeks to find (recall) the natural language question $q$.  The query length $s$ is selected with probability $P(s)$.  Then, a keyword query $k=(t_1,\dots,t_s)$ is constructed by sampling $s$ terms from $P(t_i|\theta_q)$, which is the model of $q$.  The prior probability distribution $P(s)$ can be easily estimated by considering query lengths in a representative sample (e.g., a query log).  The quality of the synthetic queries crucially depends on the distribution $P(t_i|\theta_q)$, as it determines which terms will be sampled. 
\citet{Azzopardi:2007:BSQ} define $P(t_i|\theta_q)$ using the standard language modeling approach: 
\begin{equation}
	P(t_i|\theta_q) = (1-\lambda) P(t_i|q) + \lambda P(t_i) ~.
	\label{eq:Ptithetad}
\end{equation}
Accordingly, term generation is a mixture between sampling a term from the given item with probability $P(t_i|q)$, and from the corpus with probability $P(t_i)$, where the influence of the collection model is controlled by the smoothing parameter $\lambda$. The latter likelihood is calculated using: 
\begin{equation}
	P(t_i) = \frac{n(t_i)}{\sum_{t_j {\in} V}{n(t_j)}} ~,
	\label{eq:Pti}
\end{equation}
where $n(t_i)$ denotes the collection term frequency of term $t_i$, and $V$ is the vocabulary of terms in the corpus.  

To simulate different types of user querying behavior, three plausible term selection strategies have been proposed to estimate $P(t_i|q)$: (i) popular selection, (ii) discriminative selection, and (iii) their combination~\cite{Azzopardi:2006:ACK, Azzopardi:2007:BSQ}.

(i)~\emph{Popular}: Assuming that more frequent terms are more likely to be used as query terms, $P(t_i|q)$ is calculated by Eq.~\eqref{eq:popular}, where $n(t_i, q)$ is the number of occurrences of $t_i$ in $q$.
\begin{equation}
	P(t_i|q) = \frac{n(t_i, q)}{\sum_{t_j {\in} q}{n(t_j, q)}}
	\label{eq:popular}
\end{equation}

(ii)~\emph{Discriminative}: Assuming that the user may select query terms that can better discriminate the item she is looking for from other items in the corpus, $P(t_i|q)$ is calculated using Eq.~\eqref{eq:discr}, where $b(t_i, q)$ is a binary indicator function that is $1$ if $t_i$ occurs in $q$ and $0$ otherwise. $P(t_i)$ is the same as before, cf. Eq.~\eqref{eq:Pti}.
\begin{equation}
	P(t_i|q) = \frac{b(t_i, q)}{P(t_i) \sum_{t_j {\in} q}{\frac{b(t_j, q)}{P(t_j)}}}
	\label{eq:discr}
\end{equation}

(iii)~\emph{Combination}: Combining the popular and discriminative strategies into a single model, $P(t_i|q)$ is calculated by Eq.~\eqref{eq:comb}, where $df(t_i)$ is the document (here: question) frequency of term $t_i$ and $N$ is the total number of items in the corpus. 
\begin{equation}
	P(t_i|q) = \frac{n(t_i, q) \log{\frac{N}{df(t_i)}}}{\sum_{t_j {\in} q} (n(t_j, q) \log{\frac{N}{df(t_j)}} )}
	\label{eq:comb}
\end{equation}

\subsubsection{Our Keyword Generation Algorithm}
\label{sec:generation:kqgm:modify}

Note that the original algorithm in~\citep{Azzopardi:2007:BSQ} has been developed for known-item (document) search. We need to modify and extend it at several points to be able to use it for the K2Q task we are addressing. 

For known-item search, an item $q$ is selected randomly from the corpus, and then a keyword query is generated from that item.  The process is repeated as many times as the number of queries to be created.
In our problem scenario the items are natural language questions, where each of them needs to be paired with a keyword query.  That is, we do not sample items, but we generate a query for each item in the corpus.  This is the first modification we make to the algorithm (line 3 in Algorithm \ref{alg:simul}).  

The second change concerns the length of keyword queries.
In~\citep{Azzopardi:2007:BSQ}, the length of the query is drawn  from a Poisson distribution, with the mean set according to the average length in a set of  human-generated queries.  For us, the length of the keyword query also depends on the corresponding natural language question.  Given a question with length $|q|$, it is reasonable to assume that users will always prefer to issue a keyword query that is shorter than $|q|$.   Thus, we include this additional constraint and sample a query length with $P(s)$, where $s < |q|$ (line 5 in Algorithm \ref{alg:simul}).

Third, keyword queries typically do not contain question words, such as ``how,'' ``what,'' ``where,'' ``who,'' ``why,'' ``when,'' etc. 
Thus, we do not sample question words in our generation process.  

Fourth, our algorithm does not only sample terms but also samples phrases for generating synthetic queries.  Thus, we avoid breaking up word sequences that function together as a meaningful unit.  It means that $t_i$ could be either a term or a phrase in the generative process (line 7 in Algorithm~\ref{alg:simul}).  We describe our phrase detection mechanism in \S\ref{sec:generation:kqgm:phrase}.

Fifth, according to our statistics on a sample of queries,\footnote{The Yahoo! L16 Webscope Dataset, which contains many real keyword queries from users of Yahoo Answers.} only 3.9\% of all keyword queries include the same term more than once, suggesting that queries with repeated terms are atypical.  Thus, we find it reasonable to avoid sampling the same term more than once in our keyword query generation process (line 9 in Algorithm \ref{alg:simul}).

\begin{algorithm}[t]
\SetAlgoLined
\SetKwFunction{Range}{range}
\KwData{$\mathcal{Q}$, a set of known questions}
\KwResult{${\langle}\mathcal{K},\mathcal{Q}{\rangle}$, a set of synthetic keyword-question pairs}
\Begin{
${\langle}\mathcal{K},\mathcal{Q}{\rangle} \leftarrow \emptyset$\;
 \For{$q\in \mathcal{Q}$}{
 $k \leftarrow []$\; 
 $s \leftarrow \mathrm{sampleQueryLength}(P(s))$\; 
 \For{$j$ in [1, $s$], $s < |q|$}{  
  $t_i \leftarrow \mathrm{sampleTerm}(P(t_i|\theta_q))$\;
  $k' \leftarrow \mathrm{append}(k,t_i)$\; 
  $P(t_i|\theta_q) \leftarrow 0 $\; 
  }
  ${\langle}\mathcal{K},\mathcal{Q}{\rangle} \leftarrow {\langle}\mathcal{K},\mathcal{Q}{\rangle} \cup \{{\langle}k,q{\rangle} \}$\;
 }
}
\caption{Synthetic keyword-question generation}
\label{alg:simul}
\end{algorithm}

\subsubsection{Paraphrase-Based Querying Model}
\label{sec:generation:kqgm:pqm}

Users may use different words to express the same meaning.  This should be taken into consideration in the keyword query generation process.
Imagine the following case, where a particular user has seen the question \emph{``Who is the author of the pooh?"} in a community question answering forum (e.g., Yahoo! Answers or Quora), then, after several days, she tries to recall the search terms to find an answer to this question.  If she still remembers the exact words from the question, she may issue \emph{``the pooh author"} as a query.  Otherwise, she may recollect a paraphrase of the question, like \emph{``Who is winnie the pooh's creator?"} and, based on that, formulates the keyword query \emph{``winnie the pooh creator}."
Furthermore, different users may recall different paraphrases during their querying process.  Thus, it is natural to sample terms from paraphrases of the same question when generating keyword queries.  We realize this idea by defining the term generation model $P(t_i|\theta_q)$ as a three component mixture: 
\begin{equation}
	P(t_i|\theta_q) = \alpha P(t_i|q) + \beta P(t_i| C_{q}) + (1 - \alpha - \beta) P(t_i) ~,
	\label{eq:paraphrase}
\end{equation}
where $C_{q}$ is a set of paraphrases of question $q$ and $P(t_i|C_{q})$ defines the likelihood of selecting term $t_i$ from the paraphrases.  All paraphrases in $C_{q}$ are concatenated together into a single large document, then $P(t_i|C_{q})$ may be calculated by one of three strategies we described in the previous section.
The model in Eq.~\eqref{eq:paraphrase} has two parameters, $\alpha, \beta \in [0, 1]$.  As $\alpha$ tends to one, it assumes that the user definitely remembers the terms of the original question.  As $\beta$ tends to one, it assumes that user does not recall the terms from the original question but knows how to paraphrase it. As both $\alpha$ and $\beta$ tend to zero, it means that user knows that the question exists but does not remember any terms from the original question nor from any of its paraphrases. 

\subsubsection{Phrase Detection}
\label{sec:generation:kqgm:phrase}

We sample not only terms but also phrases, in order to avoid breaking up continuous word sequences that constitute meaningful units.  Specifically, we follow the method proposed by~\citet{Mikolov:2013:DRW} for detecting phrases.  Words that belong to the same phrase are grouped together into a new term.  For example, the question \emph{``how fast is a 2004 honda crf 230"} is converted to \emph{``how fast is a 2004 honda\_crf\_230"} after phrase detection.  This way, KQGM is able to directly sample \emph{honda\_crf\_230}, instead of sampling three independent terms.

\subsection{Filtering Mechanisms}
\label{sec:generation:filter}

To ensure that high-quality synthetic data is generated for training neural translation models, we propose two filtering mechanisms.  One operates on the level of individual questions and selects the best keyword query, from a pool of candidate queries generated for a given question (\S\ref{sec:generation:filter:1}).
The other filter is applied over the entire set of synthetic query-question pairs and filters out low-quality instances (\S\ref{sec:generation:filter:2}).

\subsubsection{Keyword Query Filter}
\label{sec:generation:filter:1}

\begin{figure}[t]
\centering
\includegraphics[width=7.5cm]{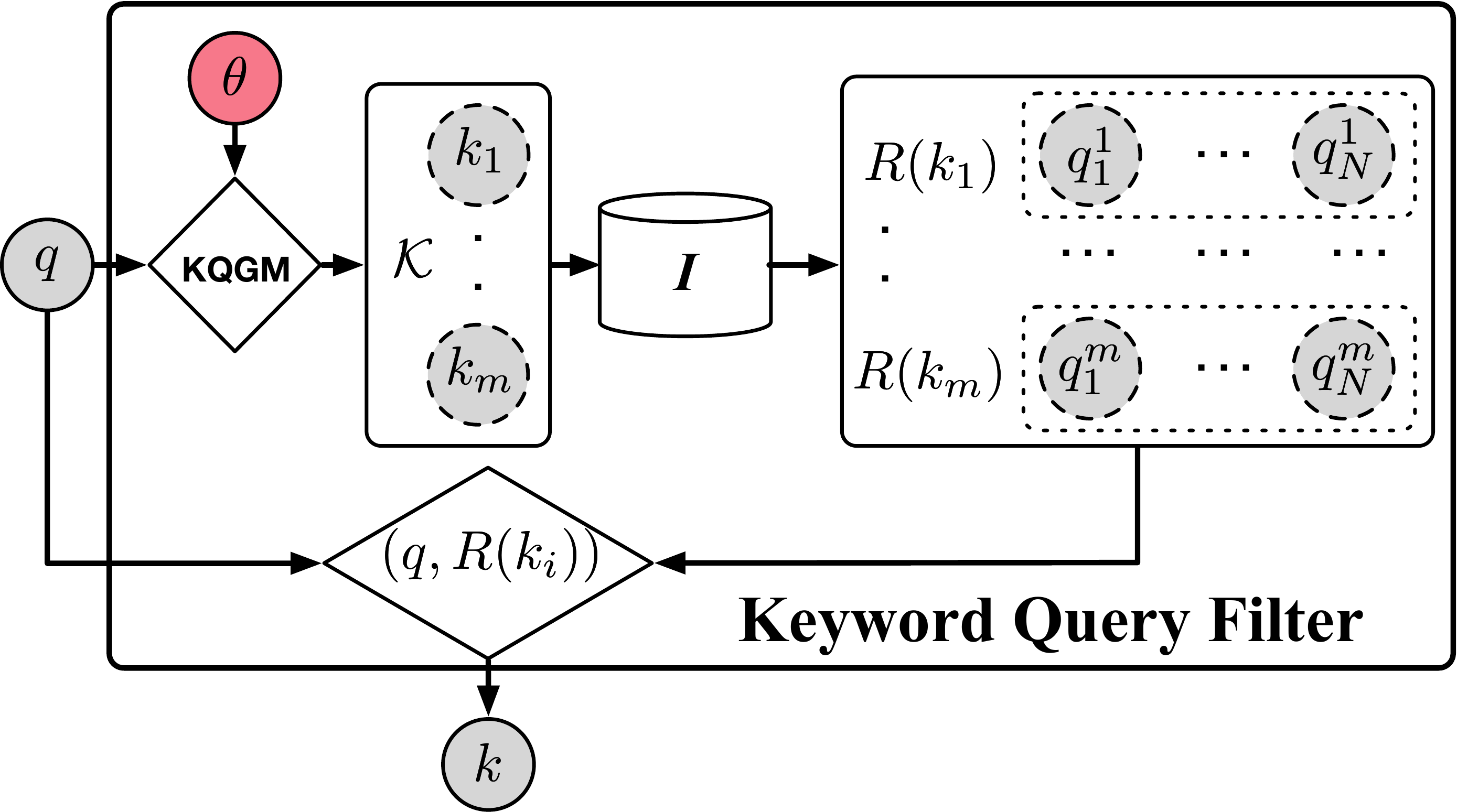}
\caption{The architecture of our keyword query filter (KQF).  For a given question $q$, $\mathcal{K}$ holds the candidate keywords generated by KQGM, out of which a a single best keyword $k$ is selected. $I$ is the index containing all questions in our corpus.  For each generated keyword $k_i \in \mathcal{K}$, $R(k_i)$ is the top-$N$ relevant questions retrieved from $I$.}
\label{fig:rf}
\end{figure}

Given the probabilistic nature of query length selection (line 5 in Algorithm~\ref{alg:simul}) and term selection (line 7 in Algorithm~\ref{alg:simul}), the keyword query generation model may produce very different keyword queries for the same question.  These keywords may vary a lot in terms of quality, from appropriate to inadequate.  For example, given the question \emph{``what happens inside a refracting telescope,"} the query generation model can give rise to a good keyword query, \emph{``happens inside refracting telescope,"} or to a rather bad one, \emph{``inside colors type,"} using the very same parameters.

The idea is to remedy this behavior by generating, for each question, a set of candidate keyword queries (i.e., running the model multiple times), and then selecting the single most suitable query.  We propose to achieve this using a so called \emph{keyword query filter} (KQF), shown in Fig.~\ref{fig:rf}.
The intuition behind this ranking-based filtering approach is that the better the generated keyword query is, the more effectively it can retrieve the original question from the question corpus.
(It is worth pointing out that our algorithm will always generate a keyword query that is shorter than the corresponding question, i.e., it is never the same as the question.)

We start with generating a set of $m$ candidate keywords $\mathcal{K} = \{ k_{1}, \dots, k_{m} \}$ for a given question $q$ using KQGM.  Then, we issue each candidate keyword query $k_{i}$ against an index containing all questions in our corpus, and retrieve the top-$N$ highest scoring questions, $R(k_i) = \langle q_{1}^{i}, q_{2}^{i}, \dots, q_{N}^{i} \rangle$.  Specifically, we employ the Sequential Dependence Model (SDM) retrieval method~\citep{Metzler:2005:MRF}.  
Finally, we select the best candidate keyword $k$ for the input question $q$ according to its reciprocal rank:
\begin{equation}	
	k = \argmax_{i \in [1 \dots m]}\frac{1}{\mathit{rank}(q, R(k_i))} ~,  \label{eq:4.2.1}
\end{equation}
where $\mathit{rank}(q, R(k_i))$ is the rank of $q$ in the ranked list $R(k_i)$.

\subsubsection{Training Data Filter}
\label{sec:generation:filter:2}

\begin{figure}[t]
\centering
\includegraphics[width=8cm]{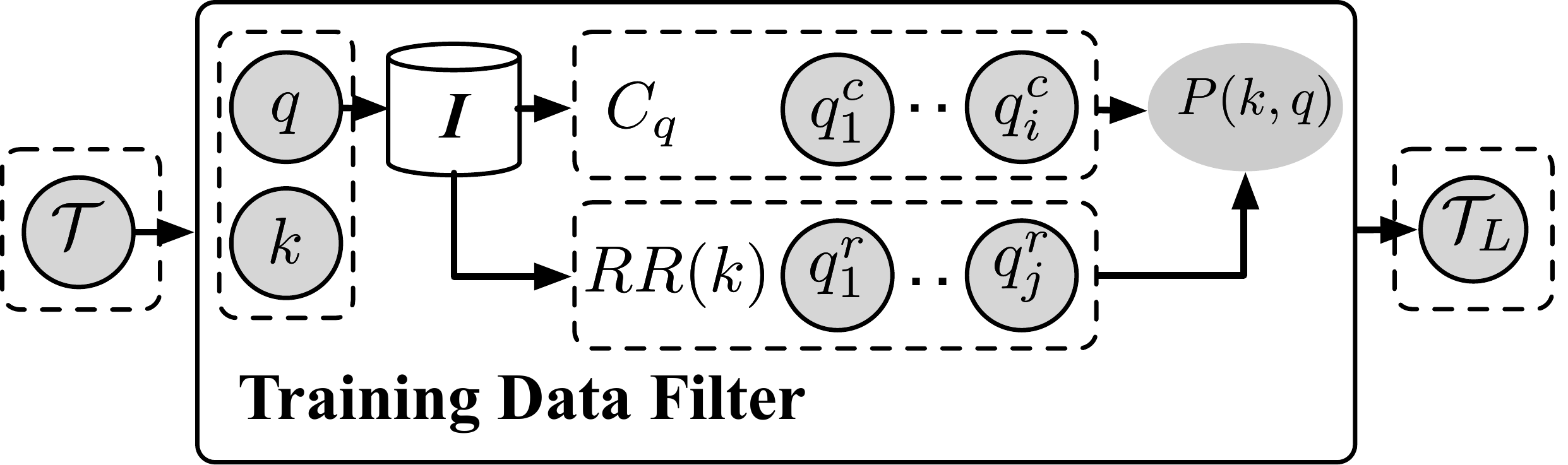}
\caption{The architecture of our training data filter (TDF).  $\mathcal{T}$ denotes the entire set of keyword-question pairs generated by KQGM with KQF.  For a pair $(k,q) \in \mathcal{T}$, $C_q$ is a set of all paraphrases of $q$ and $RR(k)$ is the set of relevant questions retrieved from the question corpus $I$ in response to $k$. $\mathcal{T}_L \subseteq \mathcal{T} $ denotes the top-$L$ pairs with highest quality score $P(k,q)$.}
\label{fig:mrf}
\end{figure}

Even after applying the keyword query filter, there may still exist low-quality training instances in $\mathcal{T}$, which would misdirect the learning process.
Therefore, we propose a \emph{training data filter} (TDF) to filter out low quality instances.  TDF, shown in Fig.~\ref{fig:mrf}, takes a set of synthetic query-question pairs $\mathcal{T}$ as input, and returns a subset $\mathcal{T}_L \subseteq \mathcal{T}$ that contains the top-$L$ pairs with the highest \emph{quality score}.  We use retrieval precision as a quality indicator, which expresses to what extent $k$ is a proper keyword for question $q$:

\begin{equation}
\begin{split}
	P(k,q) = \frac{|C_q \cap RR(k)|}{|RR(k)|} ~,
	\label{eq:6.3.1}
\end{split}
\end{equation}
where $RR(k)$ denotes the set of relevant questions retrieved by the keyword query $k$ using the SDM retrieval method~\cite{Metzler:2005:MRF}, and $C_q$ denotes the set of paraphrase questions for $q$.  
In short, TDF ranks all generated query-question pairs according to $P(k,q)$, then selects the top-$L$ highest scoring ones to form the filtered subset $\mathcal{T}_L$.
 
\section{Neural Machine Translation}
\label{sec:nmt}

Neural machine translation (NMT) aims to directly model the conditional probability $P(y_1, \dots , y_m|x_1, \dots , x_n)$ of translating a source sequence $(x_1, \dots , x_n)$ to a target sequence $(y_1, \dots , y_m)$.  Thus, it lends itself naturally to implement our K2Q task using NMT, by taking the keyword query as the source sequence $(x_1, \dots , x_n)$ and the natural language question as the target sequence $(y_1, \dots , y_m)$.  In the rest of this section, we detail three NMT networks we use in our experiments. 

\subsection{Encoder-Decoder NMT}
The classical architecture of NMT is Encoder-Decoder recurrent neural networks (RNNs)~\cite{Sutskever:2014:SSL}, which consists of two components:

(i) \emph{Encoder}, a RNN to compute a context vector representation $c$ for the input sequence, $\mathbf{x} = (x_1, \dots , x_n)$, by iterating the following equations:
\begin{align}
    h_{i} = f(x_{i}, h_{i-1}), ~ i \in [1 \dots n] \label{eq:4.1.1} \\
    c = \phi(h_{1}, h_{2}, \dots, h_{n}) ~, \label{eq:4.1.2}
\end{align}
where $x_{i} \in \mathbb{R}^{v}$ is a one-hot representation of the $i$th word in the input sequence, and   $h_{i} \in \mathbb{R}^{n}$ is the hidden state vector of encoder RNN at time $i$. The activation function $f$ may be as simple as a sigmoid function or complex, e.g., a long short-term memory (LSTM)~\cite{Hochreiter:1997:LSM} or Gated Recurrent (GRU)~\cite{Chung:2014:GCB} unit.  
The context vector $c$ is defined by $\phi$, which is an operation on all hidden states.  In this paper, $\phi$ indicates an operation choosing the last hidden state $h_{n}$.

(ii) \emph{Decoder}, another RNN to decompress the context vector $c$ and output the target sequence, $\mathbf{y} = (y_1, \dots , y_m)$, through a conditional language model:
\begin{align}
    {s}_{i} = {f'}({s}_{i-1}, y_{i-1}, c), ~ i \in [1 \dots m] \label{eq:4.1.3} \\
	P(y_i| y_1, \dots , y_{i-1}, \mathbf{x}) = g({s}_{i}, {y}_{i-1}, c) ~, \label{eq:4.1.5}
\end{align}
where ${y}_{i} \in \mathbb{R}^{n}$ is a one-hot representation of the $i$th word in the output sequence;  $s_{i}$ denotes the hidden state vector of the decoder RNN at time $i$; $f'$ can be the same as encoder activation function, $f$, or a different non-linear activation function;  $g$ is a softmax classifier.  Given a set of keyword-question pairs, the encoder and decoder are jointly trained to maximize the conditional log-likelihood.

\subsection{Attention Mechanism}
The attention mechanism in neural networks has a long history in computer vision~\citep{Itti:1998:MSV, Paletta:2005:QSA, Mnih:2014:RMV}, and has recently been also successfully applied in natural language processing~\citep{Bahdanau:2014:NMT, Yin:2016:ACN}.  The basic idea behind it is that humans pay attention to specific parts, rather than the whole input, when performing visual and linguistic tasks.  
The attentional NMT~\citep{Bahdanau:2014:NMT} uses a dynamically changing context vector $c_i$ instead of a fixed context vector $c$ during the decoding process.  The dynamically changing context vector $c_i$ is computed with a weighted sum of the source hidden states according to: 
\begin{eqnarray}
	c_i = \sum_{j=1}^{n} {\alpha}_{ij} h_{j} & \text{and} & {\alpha}_{ij} = \frac{exp(\sigma(h_{i-1} , h_{j}))}{\sum_{j'} exp(\sigma(h_{i-1}, h_{j'}))} ~, 
	\label{eq:5.2.1}
\end{eqnarray}
where $\sigma$ is an attention function that scores the corresponding attentional strength.  Usually, $\sigma$ is parameterized with a feedforward neural network.  Further, $h_{j}$ denotes the hidden state of the encoder at time $j$, and $\alpha_{ij}$ denotes the attentional strength that the target word $y_i$ is related to a source word $x_j$.

\subsection{Copying Mechanism}
The copying mechanism was first proposed by \citet{Gu:2016:ICM} for handling out-of-vocabulary words, by selecting appropriate words from the input text.  We employ the copying mechanism to assign higher probability to words that appear in the input text.  This way we naturally capture the fact that questions tend to keep important words from the keyword query. 
By incorporating the copying mechanism into NMT, the probability of generating word $y_i$ in the output sequence becomes: 

\begin{equation}
\begin{split}
	P(y_i| y_1, \dots , y_{i-1}, \mathbf{x}) = &~~ P_{g}(y_i| y_1, \dots , y_{i-1}, \mathbf{x}) \\ 
	&  + P_{c}(y_i| y_1, \dots , y_{i-1}, \mathbf{x}) ~.
	\label{eq:5.2.2} 
\end{split}
\end{equation}
The first part is the probability of generating the term $y_i$ from vocabulary (cf. Eq.~\eqref{eq:4.1.5}). The second component is the probability of copying it from the source sequence:

\begin{align}
    P_{c}(y_i| y_1, \dots , y_{i-1}, \mathbf{x}) = \frac{exp ( {\phi}_{c}(x_j) )}{\sum_{j:x_j=y_i} {exp ( {\phi}_{c}(x_j) ) }} ~, ~ x_{j} \in \mathcal{X} \\
    {\phi}_{c}(x_j) = \sigma(h^{T}_{j}W_{c})s_{t}
\end{align}
where $\mathcal{X}$ denotes all words in the source sequence. $\sigma$ is a non-linear function and $W_{c} \in R$ is a learned parameter matrix.
We refer to~\citep{Gu:2016:ICM} for further details.
\section{Data}
\label{sec:data}

Our approach needs a small set of hand-labeled keyword-question pairs and a large set of questions.  We obtain these two datasets from WikiAnswers.\footnote{\url{http://knowitall.cs.washington.edu/oqa/data/wikianswers/}}  WikiAnswers includes millions of questions asked by humans.  Users have also identified groups of questions that are paraphrases of each other. These groups are considered paraphrase clusters~\citep{Fader:2014:OQA}.

\begin{figure}[t]
	\centering
	\begin{tabular}{cc}
		\subfigure[Question length (WikiAnswers)]{
			\includegraphics[width=5cm, height=3.8cm]{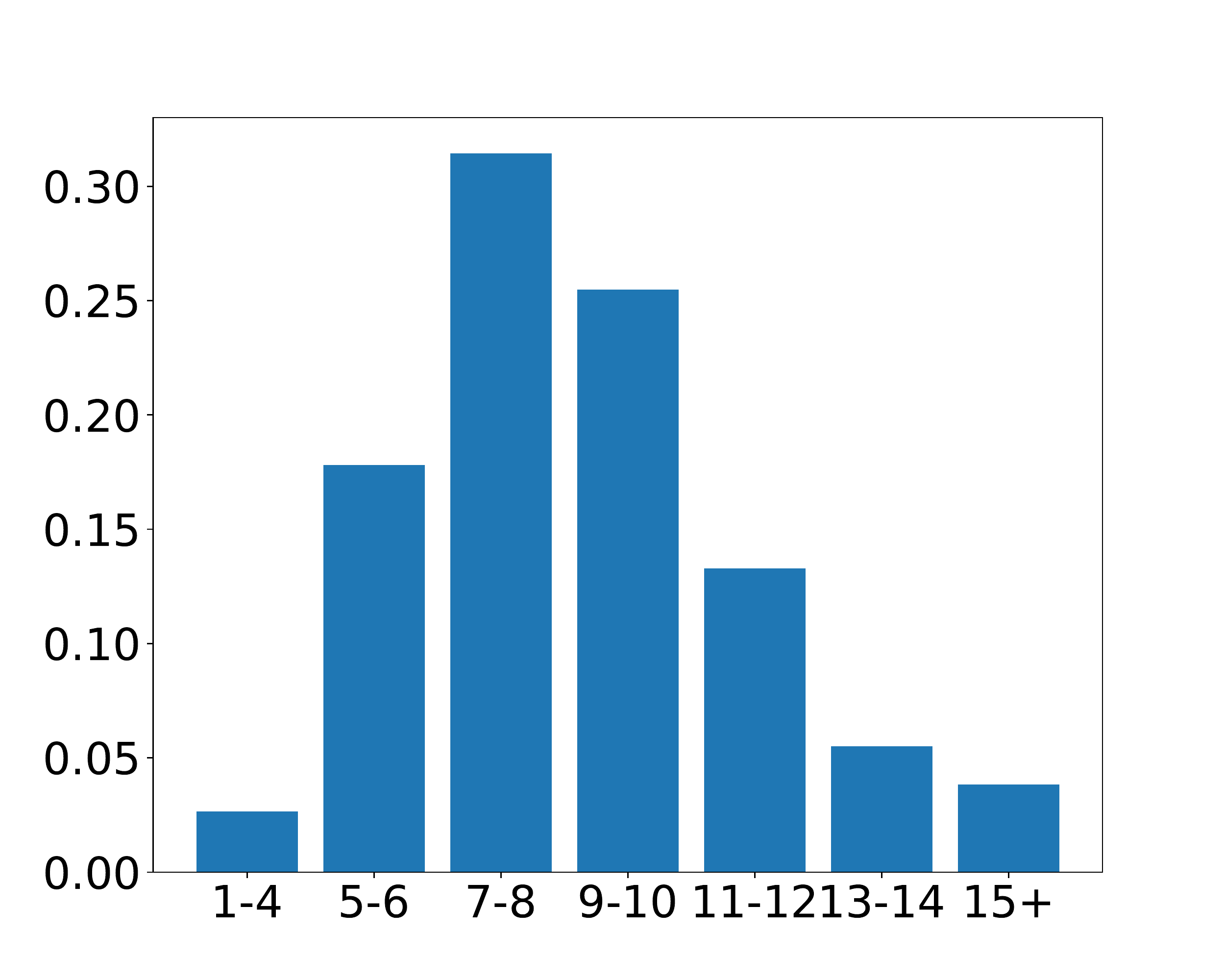}
			\label{fig:qkld:q}
		}	
		&
		\subfigure[Keyword query length (Yahoo! Webscope L16  Dataset)]{
			\includegraphics[width=3cm, height=3.8cm]{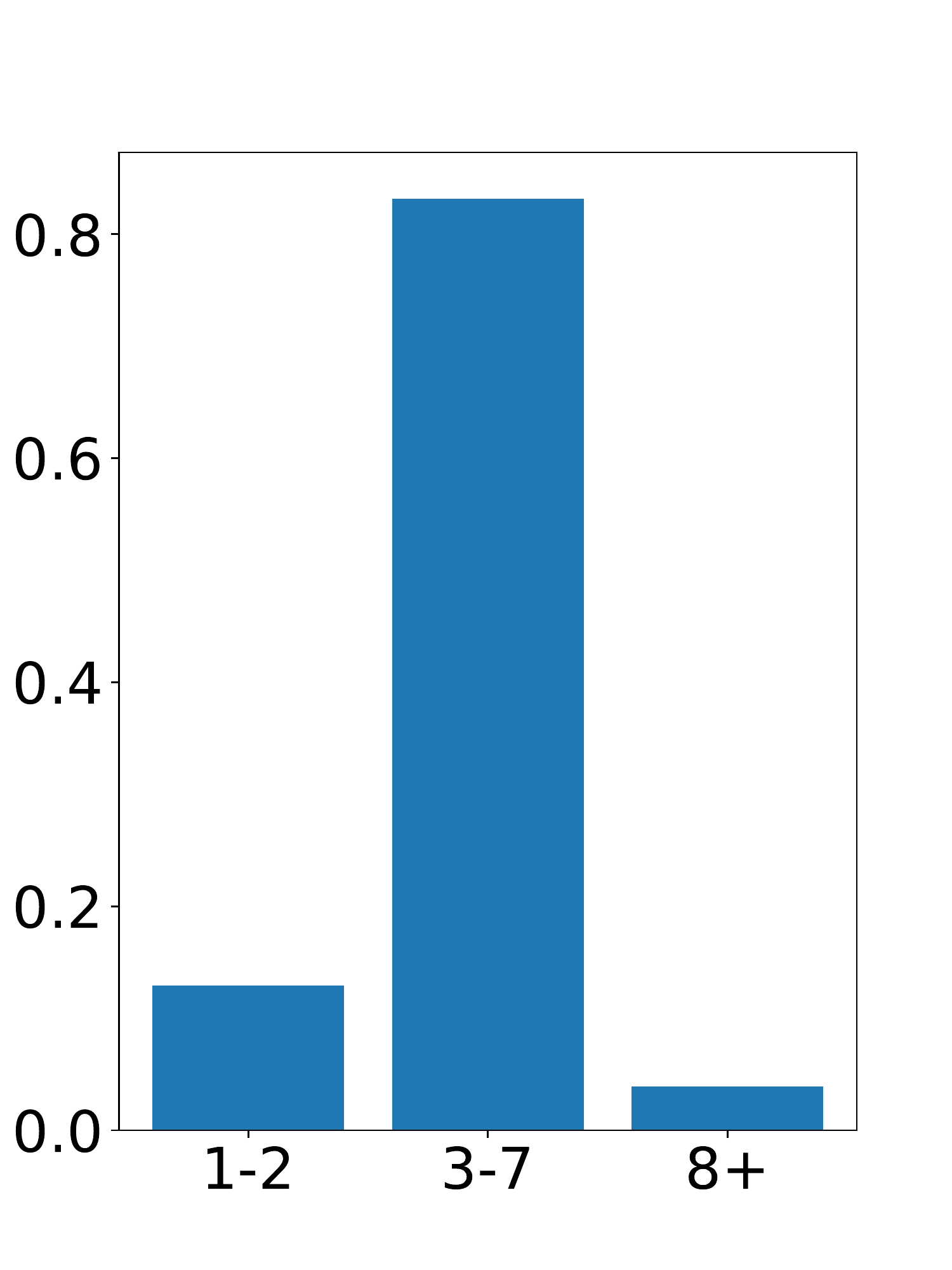}
			\label{fig:qkld:k}
			}
	\end{tabular}
	\vspace*{-0.5\baselineskip}
	\caption{Question and query length distributions. The X axes indicate length, the Y axes indicate the proportion.}
	\label{fig:qkld}
	\vspace*{-0.5\baselineskip}
\end{figure}

\paragraph{Preprocessing}
Since we only care about natural language questions in this work, we employ the heuristics proposed by~\citet{Dror:2013:QQO} to filter out non-natural language questions.  Specifically, we keep only questions that start with ``WH words" or auxiliary verbs.  
Additionally, we restrict ourselves to questions consisting of 5-12 terms (most frequent query length), based on question length distribution statistics of WikiAnswers, see Fig.~\ref{fig:qkld:q}.
We end up with 3,168,878 paraphrase clusters, with 26.05 questions per cluster on average.
In the remainder of the paper, when we write WikiAnswers, we refer to this preprocessed subset of the collection.

\paragraph{Small Set of Keyword-Question Pairs ($\mathcal{T}_0$)}
In order to get the small set of hand-labeled keyword-question pairs, we randomly pick 200 clusters from the 3,168,878 paraphrase clusters.  From each of those paraphrase clusters, we sample five questions randomly. 
We employ five human annotators, who each receive only one question from each of the 200 paraphrase clusters. The annotators then manually create keyword queries from their questions.
We then have 200 paraphrase clusters, each with five questions' paraphrases and corresponding keyword queries (where each paraphrase is labeled by a different annotator), a total of 1000 hand-labeled pairs.

\paragraph{Large Set of Questions ($\mathcal{Q}$)}
To get the large set of questions, we randomly sample a single question from each of the remaining paraphrase clusters. 
This amounts to 3,168,678 questions.   
The hand-labeled questions do not appear in this set.

\section{Experimental Setup}
\label{sec:eval:setup}

This section details various settings of three main components used in our approach, i.e., KQGM (Sect.~\ref{sec:eval:setup:kqgm}), filtering mechanisms (Sect.~\ref{sec:eval:setup:filter}), and NMT (Sect.~\ref{sec:eval:setup:nn}).

\subsection{Keyword Query Generation Model}
\label{sec:eval:setup:kqgm}

The following settings are used in our experiments:

\begin{itemize}[leftmargin=*]
	\item \emph{Query length}: The prior probability of query length $P(s)$ is calculated based on the small set of (hand-labeled) keyword-question pairs.  According to statistics on user keyword queries from the Yahoo! L16 Webscope Dataset, most keyword queries contain between 3 and 7 terms, see Fig.~\ref{fig:qkld:k}.  Thus, we only sample queries with length $s \in [3, 7]$.
	\item \emph{Collection Language Model}: The collection language model probability of $P(t_i)$ is computed based on the WikiAnswers dataset.  For the paraphrase-based model, we need to know the paraphrases $C_{q}$ for a given question $q$.  In our dataset, this is readily available.  We note that there also exist methods to detect paraphrases automatically~\cite{Bogdanova:2015:DSE,Jiang:2017:SQM}.
	\item \emph{Parameter Tuning}: For the baseline model (\S\ref{sec:generation:kqgm:bl}), there is only one free parameter $\lambda \in [0.1, 0.9]$.   The paraphrase-based model (\S\ref{sec:generation:kqgm:pqm}) involves two parameters, $\alpha \in [0.1, 0.9]$ and $\beta \in [0.1, 1-\alpha]$.  We set the parameter values by performing an extensive (grid) search in steps of $0.1$.
\end{itemize}

\subsection{Filtering Mechanisms}
\label{sec:eval:setup:filter}
For our filters, we use the following settings:

\begin{itemize}[leftmargin=*]
	\item \emph{Keyword query filter} (\S\ref{sec:generation:filter:1}): We generate $m=20$ candidate keywords for each question $q$ in the large set of questions using KQGM.  The best of these is selected by KQF to be paired with $q$.
	\item \emph{Training data filter} (\S\ref{sec:generation:filter:2}): For a keyword-question pair $(k,q)$ we retrieve the top $N=100$ questions using $k$ and obtain the paraphrases $C_q$ from paraphrase cluster of $q$.
\end{itemize}

\subsection{Neural Networks}
\label{sec:eval:setup:nn}
We implement the following three networks:
\begin{itemize}[leftmargin=*]
	\item \emph{EDNet}: Basic encoder-decoder NMT network.
	\item \emph{AttNet}: EDNet with attention mechanism.
	\item \emph{CopyNet}: AttNet plus copying mechanisms.
\end{itemize}
For all three networks, we choose the top 44K most frequent words in WikiAnswers as our vocabulary.  We set the embedding dimension to 100, and initialize the word embeddings randomly with a uniform distribution in [-0.1,0.1].  We set the number of layers of both encoder and decoder RNNs to 1.  Further, we use a bidirectional GRU~\cite{Bahdanau:2014:NMT} unit with size 200 for encoder RNNs, and a GRU unit with size 400 for decoder RNNs.  All networks are optimized using Adam~\cite{Kingma:2014:AMS} with an initial learning rate of $10^{-4}$, gradient clipping of $0.1$, and dropout rate of $0.5$.

\subsection{Preliminary Study}
\label{sec:eval:pre}

Our synthetic data generation heavily depends on the generative model for creating keyword queries.  Thus, we perform a preliminary study, using the small set of keyword-question pairs, $\mathcal{T}_0$, to analyze the performance of various KQGM configurations.  Informed by this analysis, we can decide which of the three term selection strategies to use for KQGM in our main experiments.

\subsubsection{Evaluation Metrics.}
\label{sec:eval:pre:metric}
We use automatic metrics from text summarization, specifically, the widely used ROUGE-L metric~\citep{Lin:2004:RPA}.
ROUGE-L not only awards credit to in-sequence unigram matches, but also captures word order in a natural way.  Thus, it can effectively measure the degree of match between the synthetic and ground truth keyword queries.
Recall that in our dataset, we have a set of paraphrases for each question.  We wish to consider those paraphrases as well in our evaluation.
Formally, let $k$ denote the generated keyword query corresponding to question $q$; $C_q$ denotes the paraphrase cluster of $q$; $\mathcal{K}_q$ is the set of ground truth keywords corresponding to $C_q$. 
For scoring $k$, we consider the set of ground truth keywords $\mathcal{K}_q$ in two different ways: 
(i) by computing the average ROUGE-L between $k$ and each ground truth keyword $k' \in \mathcal{K}_q$ (Eq.~\eqref{eq:avgrougel}), and
(ii) by considering only the best (highest scoring) ground truth keyword query (Eq.~\eqref{eq:maxrougel}).

\begin{equation}
\begin{split}
	AvgRougeL = \frac{\sum_{k' \in \mathcal{K}_q}{{\mathit{RougeL}(k, k')}}}{|\mathcal{K}_q|} 
	\label{eq:avgrougel}
\end{split}
\end{equation}
\begin{equation}
\begin{split}
	MaxRougeL = \max_{k' \in \mathcal{K}_q}{\mathit{RougeL}(k, k')} 
	\label{eq:maxrougel}
\end{split}
\end{equation}
We employ five-fold cross-validation for evaluation.
To eliminate the effects of randomness that is involved in the process, we repeat 100 times, and report the means and standard deviations.

\subsubsection{Summary.}
\label{sec:eval:pre:summary}
Table~\ref{tbl:res_q2k} shows the evaluation results for all KQGM configurations.  Comparing the three term selection strategies (\S\ref{sec:generation:kqgm:bl}), we find that the \emph{Combination} strategy always attains the best performance.  With the same term selection strategy and KGQM, phrase detection brings noticeable improvements in both AvgRougeL and MaxRougeL (+5.28\% and +4.22\%, respectively).  Comparing the paraphrase-based model with the baseline model, the former brings +10.66\% improvements on average for AvgRougeL and +7.16\% on average for MaxRougeL.  The paraphrase-based model with phrase detection achieves the best overall performance, with 0.2521 AvgRougeL and 0.3843 MaxRougeL, which is superior to the best baseline configuration.
\begin{table}[t]
\centering
\caption{Evaluation of various KQGM configurations.  All numbers are obtained using 5-fold cross-validation.  In parentheses are the standard deviations.}
\begin{tabular}{p{0.1cm}p{2.5cm}p{2.2cm}p{2.2cm}}
	\toprule
	\multicolumn{2}{l}{\textbf{Configuration}} & \textbf{AvgRougeL} & \textbf{MaxRougeL}\\
	\midrule
	\multicolumn{4}{l}{\emph{Baseline model}} \\
	& Popular & 0.1956 (0.0934) & 0.3197 (0.1266) \\
	& Discrimination & 0.1877 (0.1049) & 0.2999 (0.1421) \\
	& Combination & 0.2240 (0.0953) & 0.3522 (0.1331) \\
	\midrule
	\multicolumn{4}{l}{\emph{Baseline model + phrase detection}} \\
	& Popular & 0.2069 (0.1008) & 0.3354 (0.1342) \\
	& Discrimination  & 0.2062 (0.1106) & 0.3243 (0.1465) \\
	& Combination & 0.2373 (0.1019) & 0.3708 (0.1399) \\
	\midrule
	\multicolumn{4}{l}{\emph{Paraphrase-based model}} \\
	& Popular & 0.2125 (0.0930) & 0.3390 (0.1250) \\
	& Discrimination  & 0.2266 (0.1017) & 0.3458 (0.1367) \\
	& Combination & 0.2435 (0.0956) & 0.3734 (0.1330) \\
	\midrule
	\multicolumn{4}{l}{\emph{Paraphrase-based model + phrase detection}} \\
	& Popular & 0.2182 (0.1001) & 0.3476 (0.1322) \\
	& Discrimination  & 0.2355 (0.1020) & 0.3513 (0.1361) \\
	& Combination & \textbf{0.2521} (0.1009) & \textbf{0.3843} (0.1374) \\	
	\bottomrule
\end{tabular}
\label{tbl:res_q2k}
\vspace*{\baselineskip}
\end{table}
\subsubsection{Parameters.}
\label{sec:eval:pre:parameter}
\begin{figure*}[t]
	\centering
	\begin{tabular}{cc}
		\subfigure[]{
			\includegraphics[width=7.5cm]{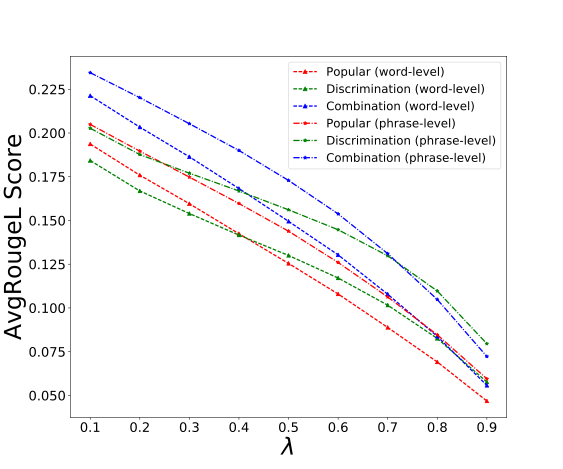}
			\label{fig:lambda_avg}
		}	
		&
		\subfigure[]{
			\includegraphics[width=7.5cm]{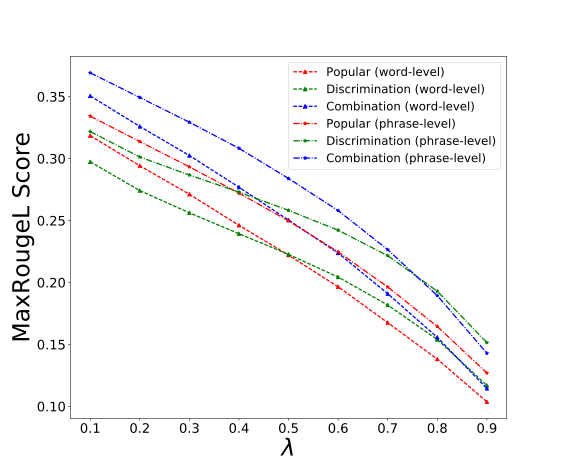}
			\label{fig:lambda_max}
			}
	\end{tabular}
	\begin{tabular}{cc}
		\subfigure[]{
			\includegraphics[width=7.5cm]{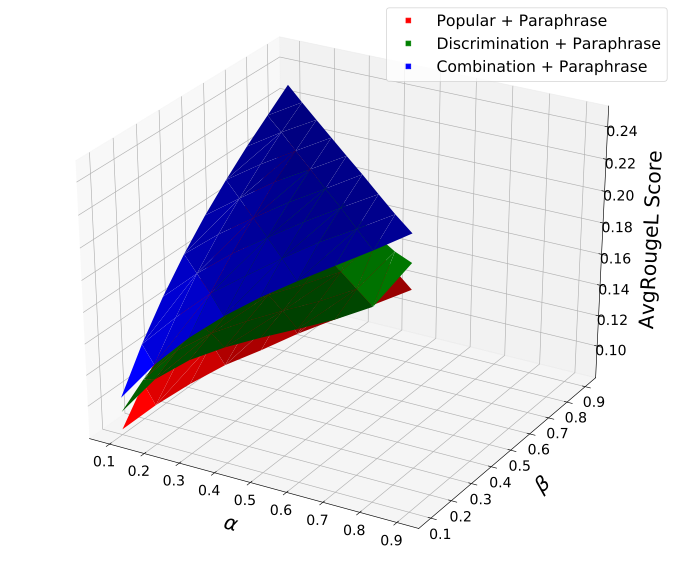}
			\label{fig:alpha_beta_avg}
		}	
		&
		\subfigure[]{
			\includegraphics[width=7.5cm]{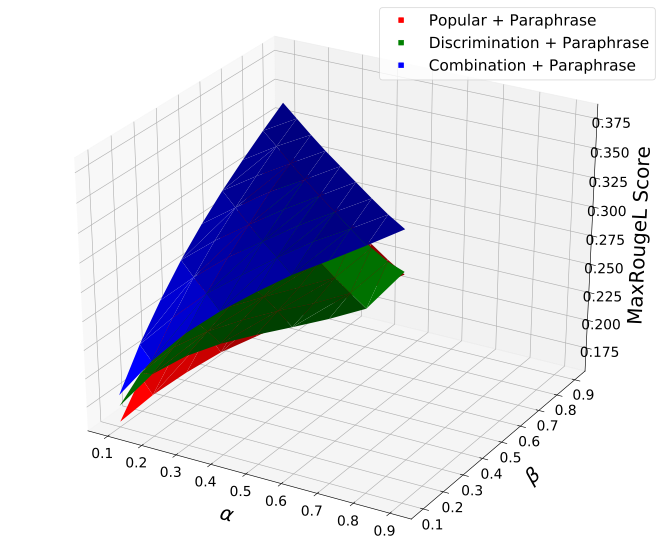}
			\label{fig:alpha_beta_max}
			}
	\end{tabular}
	\caption{Influence of free parameters $\lambda$, $\alpha$, and $\beta$ on KQGM performance.} 
	\label{fig:param_change}
\end{figure*}

We test what influence the free parameters have on the performance of KQGMs.  For the baseline model, we find that both AvgRougeL and MaxRougeL decrease as $\lambda$ increases, see Figs.~\ref{fig:lambda_avg} and~\ref{fig:lambda_max}.  For the paraphrase-based model, we find that both AvgRougeL and MaxRougeL increase with higher $\alpha$ and $\beta$ values, see Figs.~\ref{fig:alpha_beta_avg} and ~\ref{fig:alpha_beta_max}.  This is not unexpected, since users prefer to use terms from the given question for the keyword query.

\subsubsection{Observed Errors.}
\label{sec:eval:pre:error}
Based on manual inspection of synthetic keyword-question pairs, we find that the most prominent flaws in our synthetic data are extraneous terms in the KQGM-made keywords.  For example, given the question \emph{``what is usage of erw pipe,"} our KQGM generates a keyword query \emph{``erw pipe usage made meant,"} where ``made meant" are unnecessary terms.

\subsection{Implemented Systems}

\subsubsection{Baseline systems.} 
\label{sec:eval:auto:run}
We implement the SDM retrieval model \citep{Metzler:2005:MRF} and the state-of-the-art template-based method (TBM)~\citep{Dror:2013:QQO} as baselines.  The template-based K2Q method requires millions of hand-labeled keyword-question pairs from a query log, which we do not have access to.  Thus, we use our simulated keyword-question pairs instead of hand-labeled keyword-question pairs and compute term similarity using word2vec vectors, instead of TF-IDF weighted context vectors.
For the baseline systems, we retrieve the best matching question for each keyword query.

\subsubsection{Neural systems.} 
We train a neural network model with synthetic data, then feed the keyword query into the trained neural network model, to generate the most probable question.
Specifically, we use the best KQGM configuration (paraphrase-based model with combination selection strategy and phrase detection), along with the keyword query filter to generate synthetic data (a total of 3,168,678 keyword-question pairs).  Then, we use the training data filter to rank all keyword-question pairs.
\section{Experimental Results}
\label{sec:eval}

This section reports our evaluation results for the K2Q task.
First, in Sect.~\ref{sec:eval:auto}, we measure the quality of the generated questions using machine translation metrics.  Then, in Sect.~\ref{sec:eval:manual}, we employ human judges to assess a sample of questions along two dimensions: relevance and grammar. 

\subsection{Automatic Evaluation}
\label{sec:eval:auto}

We use $\mathcal{T}_0$ for the automatic evaluation of our K2Q methods, which comprises 1000 hand-labeled keyword-question pairs.
Note that these keyword-question pairs have not been used for the training of neural K2Q models.  Therefore, it is appropriate to use $\mathcal{T}_0$ as a test dataset.  We report on widely-used machine translation metrics: BLEU~\cite{Papineni:2002:BMA} and different variants of ROUGE~\cite{Lin:2004:RPA}.

\subsubsection{Results.}
\label{sec:eval:auto:summary}
Table~\ref{tbl:res_k2q} presents the evaluation result for the baseline systems and for the three neural networks.  Clearly, all NMT approaches perform better than the SDM baseline.  
As expected, the template-based method performs better than SDM, but it is still far behind CopyNet, which is the best neural method.
Compared with the basic encoder-decoder NMT network, we find that the attention mechanism brings in noticeable improvements in ROUGE-L (+13.99\%), ROUGE-1 (+9.78\%), ROUGE-2 (+16.76\%) and in BLEU (+20.59\%) scores.  Because of the extraneous terms issue (cf. \S\ref{sec:eval:pre:error}) in our synthetic data, the attention mechanism plays a very important role in skipping those terms (by assigning small weights to extraneous terms in the decoding process).  Additionally, the copying mechanism brings further minor improvements in ROUGE-L (+3.44\%), ROUGE-1 (+5.67\%), ROUGE-2 (+5.18\%) and BLEU (+1.25\%).  

\begin{table}[t]
\centering
\caption{Automatic evaluation results of baseline systems and three neural networks.  }
\begin{tabular}{l r r r r}
	\toprule
	\textbf{Method} & \textbf{ROUGE-L} &
		\textbf{ROUGE-1} &
		\textbf{ROUGE-2} &
	 	\textbf{BLEU}\\
	\midrule
	SDM & 0.3650 & 0.4123 & 0.1940 & 0.2780 \\
	TBM & 0.4357 & 0.5134 & 0.2056 & 0.2858 \\
	EDNet & 0.4338 & 0.5236 & 0.2464 & 0.3045 \\
	AttNet & 0.4945 & 0.5748 & 0.2877 & 0.3672 \\
	CopyNet & \textbf{0.5115} & \textbf{0.6074} & \textbf{0.3026} & \textbf{0.3718} \\
	\bottomrule
\end{tabular}
\label{tbl:res_k2q}
\end{table}

\subsubsection{Analysis.}
\label{sec:eval:auto:analysis}

\begin{figure}[t]
\centering
\includegraphics[width=8cm]{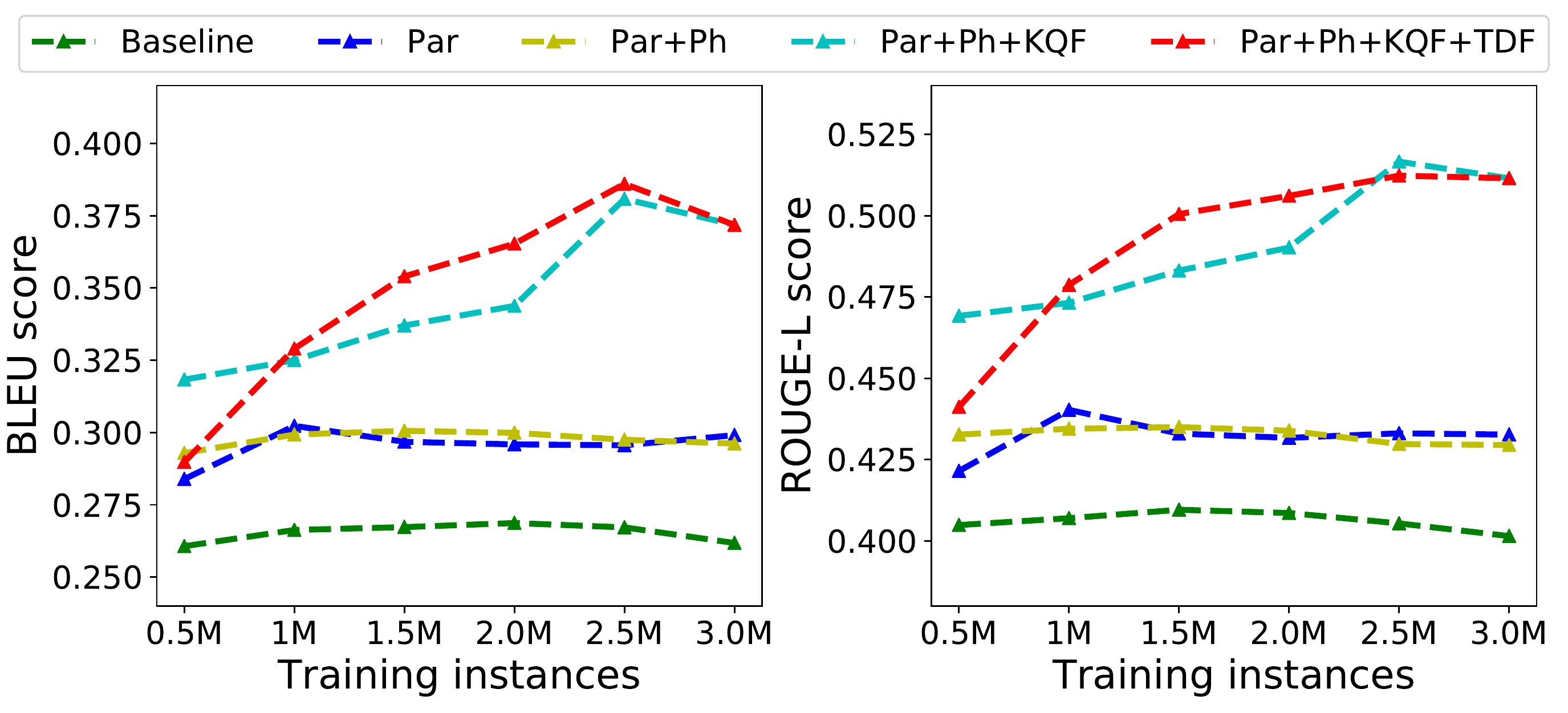}
\caption{The influence of different components of our synthetic data generation approach on the end-to-end K2Q task. The x-axis represents the amount of training data ($L$); the y-axis indicates the BLEU/ROUGE-L score.}
\label{fig:e2e}
\end{figure}

We seek to gain a better understanding of how the different elements of our synthetic data generation approach contribute to end-to-end performance on the K2Q task. For that reason, we train the best performing neural model (CopyNet) using different configurations for generating synthetic training data. We add components one by one, to see how they affect performance. 
Additionally, we vary the amount of training data used $L$ between 0.5M and 3M pairs.  The results are shown in Fig~\ref{fig:e2e}.

\begin{itemize}[leftmargin=*]
	\item \emph{Baseline}: Baseline KQGM with the \emph{Combination} term selection strategy (\S\ref{sec:generation:kqgm:bl}).
	\item \emph{Par}: Paraphrase-base KQGM with the \emph{Combination} term selection strategy (\S\ref{sec:generation:kqgm:pqm}). 
	\item \emph{Par+Ph}: Phrase detection added on top (\S\ref{sec:generation:kqgm:phrase}).
	\item \emph{Par+Ph+KQF}: Keyword query filter added on top (\S\ref{sec:generation:filter:1}).
	\item \emph{Par+Ph+KQF+TDF}: Training data filter employed on top (\S\ref{sec:generation:filter:2}).
\end{itemize}

\noindent
The first three methods do not involve the keyword query filter.  In those cases, we generate 20 candidate keyword queries for a given question and randomly select one of those.  
Only the last method uses TDF, which is a mechanism to select the top-$L$ highest quality training instances (keyword-question pairs) into $\mathcal{T}_L$.  For the other methods, we randomly select $L$ instances from the entire synthetic training data set to form $\mathcal{T}_L$.  
We run methods that involve randomization three times and report the means.  

From Fig.~\ref{fig:e2e}, we make the following observations.
First, we find the results similar to that of the KQGM evaluation in Table~\ref{tbl:res_q2k}.   Among the three KQGMs, the \emph{Par+Ph} model performs best.  The paraphrase-based KQGM brings noticeable improvements compared to baseline-based KQGM in both ROUGE-L (+6.37\% on average) and BLEU (+11.4\% on average), while adding phrase detection on top of that only brings minor improvements in ROUGE-L (+0.13\% on average) and BLEU (+0.71\% on average).

Second, comparing the results of \emph{Par+Ph} and \emph{Par+Ph+KQF}, we find that the keyword query filter brings noticeable improvements in both ROUGE-L (+13.4\% on average) and BLEU (+16.3\% on average).  Notice that by adding the keyword query filter, the performance of neural models improves with the size of the training data.  Thus, the keyword query filter is an essential element in our synthetic data generation approach.  

\begin{figure}[t]
\centering
\includegraphics[width=6cm]{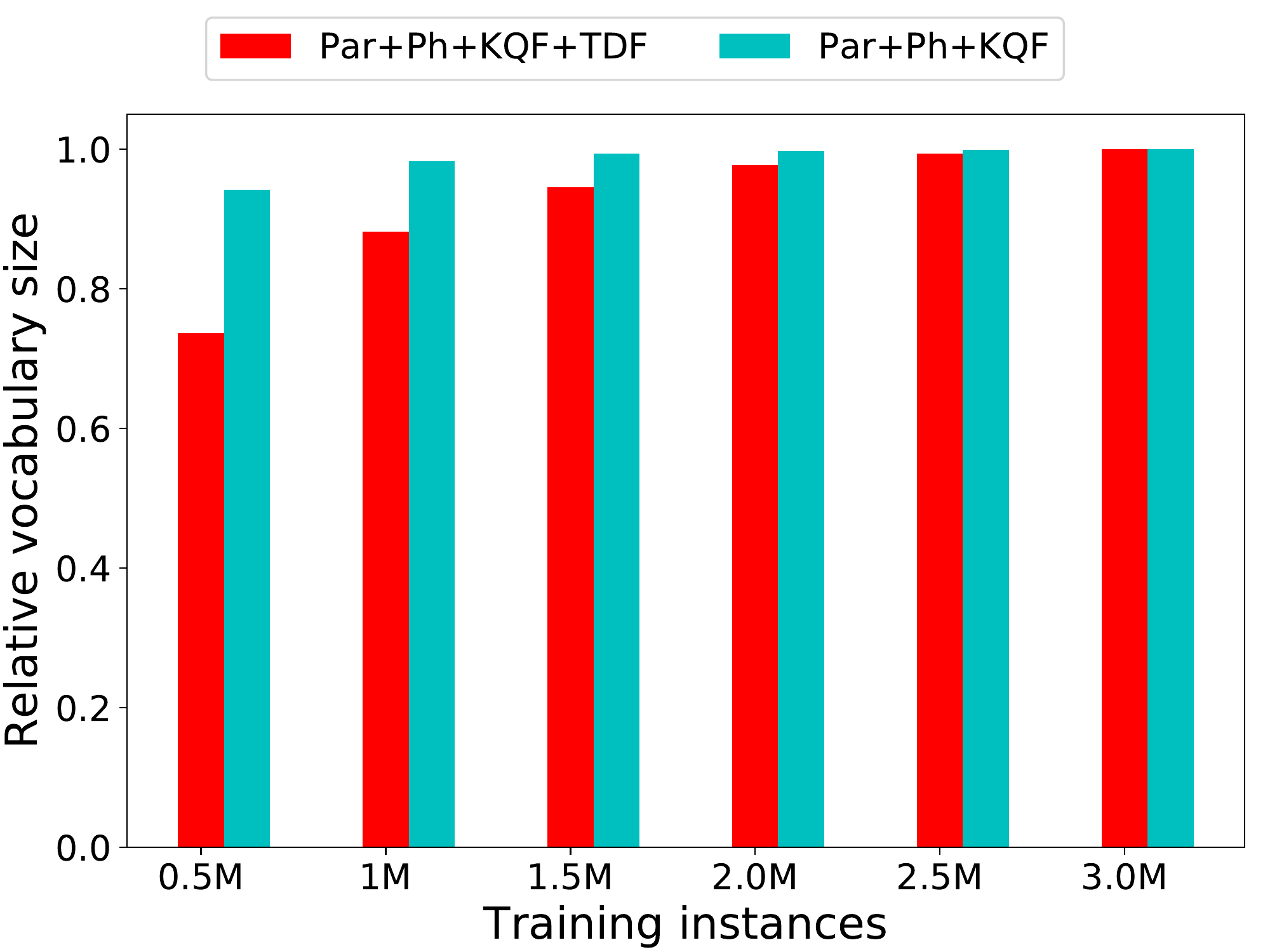}
\caption{Fraction of the total vocabulary (y-axis) captured within the subset of training instances selected by TDF (x-axis).  I.e., unique words present in $\mathcal{T}_L$, relative to $\mathcal{T}$.} 
\label{fig:vocab}
\end{figure}

Third, we find that \emph{Par+Ph+KQF+TDF} almost always performs better than \emph{Par+Ph+KQF}, demonstrating that our training data filter is able to estimate the quality of the generated keyword-question pairs, and feed high-quality training instances into the neural networks.   One noticeable exception (for both BLEU and ROUGE-L) is the leftmost data point ($L=0.5M$), where the performance of \emph{Par+Ph+KQF+TDF} is much below that of \emph{Par+Ph+KQF}.  A further analysis reveals that this is caused by an ``insufficient vocabulary'' issue.  This is illustrated on Fig.~\ref{fig:vocab}, where we plot the fraction of the total vocabulary (i.e., unique words in $\mathcal{T}$) present in the training subset $\mathcal{T}_L$.  We can observe that with only $0.5M$ training instances, the \emph{Par+Ph+KQF+TDF} model has built up only 74\% of the vocabulary, as opposed to 94\% by the \emph{Par+Ph+KQF} model.  Our training data filter, based on a retrieval method, performs well with frequent terms, but fails on rare terms.  
It appears that the TDF quality score estimator overvalues common terms and undervalues rare terms, when selecting the subset of instances $\mathcal{T}_L$ for training.

Finally, as expected from TDF, it greatly benefits performance to use the high-quality training instances first; see the \emph{Par+Ph+KQF+TDF} model for the 0.5M-1.5M range.  In contrast, the last half million training instances yield little to no improvements.  These results suggest that creating more high-quality keyword-question pairs might bring predictable improvements for neural K2Q models.

\subsection{Manual Evaluation}
\label{sec:eval:manual}

We also perform a manual evaluation using a sample of 87 real user keyword queries with low query clarity\footnote{Query clarity ranges from $1.0$ to $3.0$, where $1.0$ indicating ``clear" and $3.0$ indicating ``vague."  We only sample queries with clarity smaller than $1.5$.} from the Yahoo! Webscope L16 Dataset.  All these queries originate from the query log of Yahoo Answers.
For each keyword query, we generate 5 questions, each with a different method. That is, the SDM and TBM baselines, and the three neural networks.

\subsubsection{Assessment}
\label{sec:eval:manual:judge}

Three human raters were asked to assess each question along two dimensions: (i) \emph{Relevance}, which indicates whether the question is relevant to the keyword content-wise (ignoring grammar mistakes), and (ii) \emph{Grammar}, which reflects the grammatical correctness.  Table~\ref{tbl:rateE} shows our rating scheme.
Raters were further asked to choose the best generated question from among the five alternatives.  The number of wins were then aggregated for each of the five methods. If multiple methods generated the same question, then the point is added to all.
 
\begin{table}[t]
\centering
\caption{Human rating scheme used in our manual evaluation for relevance (R) and grammar (G).}
\begin{tabular}{p{0.1cm}p{7.5cm}}
	\toprule
	R & Rating scheme\\
	\midrule
	2 & The question is meaningful and matches given keyword\\ 
	1 & The question matches given keyword, more or less\\
	0 & The question either doesn't make sense or matches given keyword \\
	\midrule
	G & Rating scheme\\
	\midrule
	2 & No grammar errors in question, it can be understood completely\\
	1 & Few grammar errors in question, but it can be understood\\
	0 & Too many grammar errors in question, it can not be understood\\
	\bottomrule
\end{tabular}
\label{tbl:rateE}
\end{table}

\begin{table}[t]
\centering
\caption{Manual K2Q evaluation results.  The inter-rater agreement is measured using Cohen's kappa score~\cite{Cohen:1960:CAN}. Highest scores are in boldface.}
\begin{tabular}{l r r r}
	\toprule
	\textbf{Methods} & \textbf{Relevance} & \textbf{Grammar} & \textbf{Wins}\\
	\midrule
	SDM & 0.352 & \textbf{1.643} & 7.333\\
	TBM & 1.065 & 0.590 & 14.333\\
	EDNet & 0.569 & 0.682 & 6.666\\
	AttNet & 1.114 & 1.046 & 31.666\\
	CopyNet & \textbf{1.563} & 0.998 & \textbf{36.000} \\
	\midrule
	Cohen's kappa score & 0.499 & 0.498 & 0.637\\
	\bottomrule
\end{tabular}
\label{tbl:res_k2q_m}
\end{table}

Table~\ref{tbl:res_k2q_m} shows the results of human judges; the reported scores are means.  As expected, the SDM method scores highest on grammar, since it retrieves existing questions from the corpus.  However, it achieves a very low score on relevance, since it can only retrieve questions that have been asked before (i.e., exist in the corpus).  As in the automatic evaluation results, the attention mechanism brings in substantial improvements over the simple Encoder-Decoder model (both in terms of relevance and grammar).  As anticipated, the copying mechanism leads to large improvements in terms of relevance (+40.3\%); at the same time, the grammar score of CopyNet is only marginally lower than that of AttNet.

\subsubsection{Case study}
Table~\ref{tbl:ex} provides some examples of generated questions.  Clearly, SDM returns grammatically correct, but often irrelevant questions.  CopyNet has the ability to capture the meaning of the keyword query, and generates somewhat monotonous but very relevant questions.  The other two neural networks seem to capture the query intent only partially, and drift off in directions that are somewhat related to the topic of the query, yet irrelevant, e.g. ``\emph{What are cute boobs?}'' and ``\emph{What is most popular in england?}.'' 

\begin{table}[t]
\centering
\caption{Examples of generated questions from our K2Q system. The methods used to generate question are [S] SDM, [T] TBM, [E] EDNet, [A] AttNet, and [C]  CopyNet.}
\begin{tabular}{p{0.05cm}p{7.3cm}}
	\toprule
	\multicolumn{2}{l}{\textbf{Keyword 1} \emph{cute yaoi animes}} \\
	\multicolumn{2}{l}{[S] Do girls watch yaoi anime?} \\
	\multicolumn{2}{l}{[T] Is it cute when yaoi are animes?} \\
	\multicolumn{2}{l}{[E] Are there a good animes are cute?} \\
	\multicolumn{2}{l}{[A] What are cute boobs?} \\
	\multicolumn{2}{l}{[C] What are cute yaoi animes?} \\
	\midrule
	\multicolumn{2}{l}{\textbf{Keyword 2} \emph{average price movie ticket 1987}} \\
	\multicolumn{2}{l}{[S] What is the average ticket price for a super bowl ticket?} \\
	\multicolumn{2}{l}{[T] What is the average price of movie ticket 1987?} \\
	\multicolumn{2}{l}{[E] What is the average price for a 1987 ticket in 1987?} \\
	\multicolumn{2}{l}{[A] What is the average price for a movie ticket?} \\
	\multicolumn{2}{l}{[C] What is the average price of the movie ticket in 1987?} \\
	\midrule
	\multicolumn{2}{l}{\textbf{Keyword 3} \emph{popular jbs england}} \\
	\multicolumn{2}{l}{[S] How big are jbs feet?} \\
	\multicolumn{2}{l}{[T] Who are popular sovereignty and jbs england related?} \\
	\multicolumn{2}{l}{[E] What is the most popular in england?} \\
	\multicolumn{2}{l}{[A] Who is popular in england?} \\
	\multicolumn{2}{l}{[C] How popular is jbs in england?} \\
	\bottomrule
\end{tabular}
\label{tbl:ex}
\end{table}

\subsubsection{Comparison to Google}
\begin{figure}[t]
\centering
\includegraphics[width=8cm]{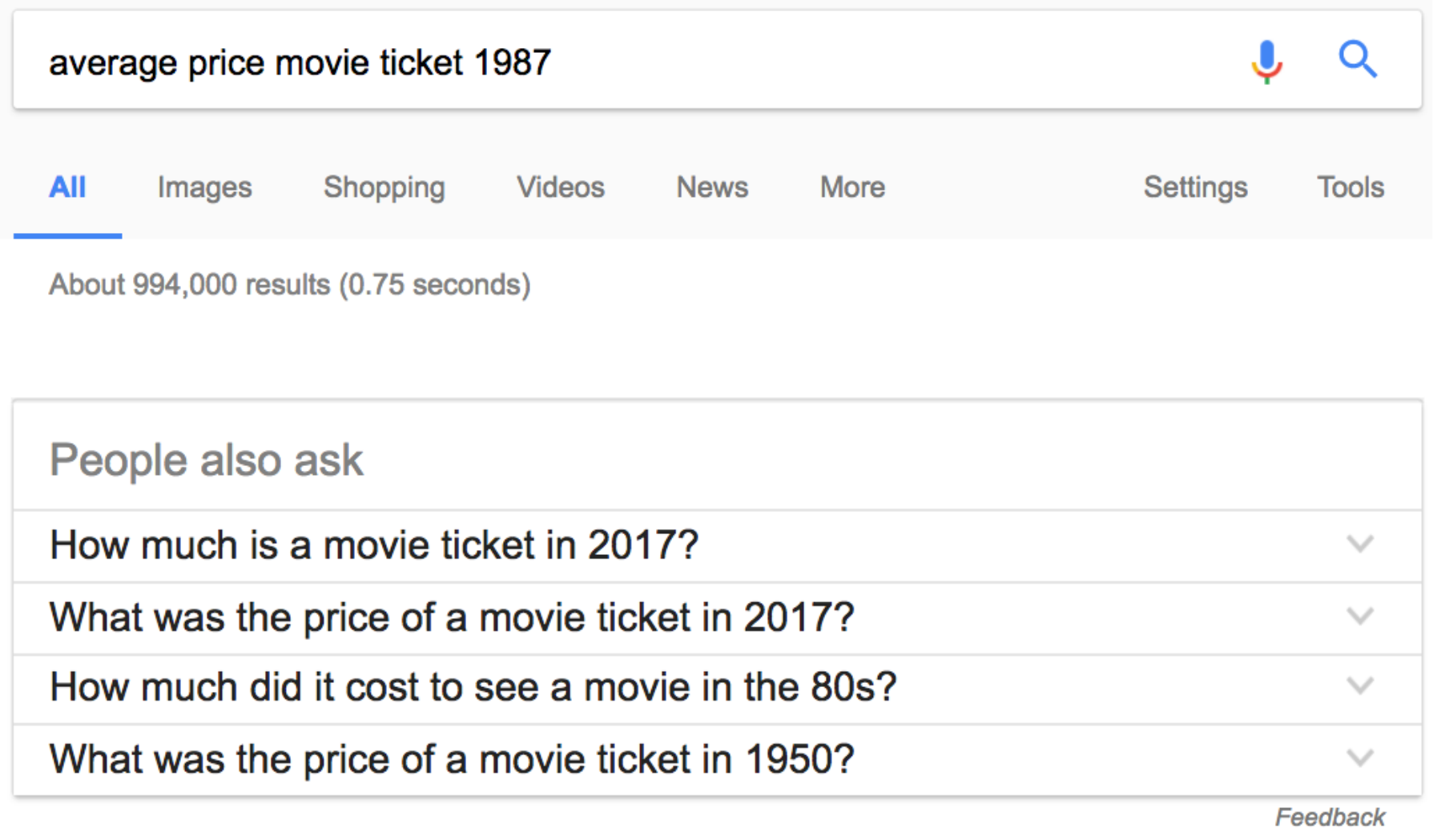}
\vspace*{-0.5\baselineskip}
\caption{Screenshot of Google's ``People also ask'' service (captured on January 15, 2018) for the query ``average price movie ticket 1987.''}
\label{fig:googlex}
\vspace*{-\baselineskip}
\end{figure}

The ``People also ask'' service, provided by Google and illustrated on Fig.~\ref{fig:googlex}, is somewhat similar to our keyword-to-question task.  Therefore, we considered including it in our evaluation for baseline comparison.  When running our test queries through Google, we noticed that the ``People also ask'' panel is triggered only for 34\% of our queries.  
As we already pointed out in the introduction, for each query, we wish to generate a natural language questions that most likely represents the user's underlying information need.  It appears that Google's service is addressing a fundamentally different task, which is of suggesting questions related to the query that are asked by sufficiently many people and to which answers are known to exist.
Because of these, we do not compare our methods against this service.

\section{Related work}
\label{sec:related}
In this section we review related research from two areas: keyword-to-question and synthetic data generation.

\subsection{Keyword to Question}
\label{subsec:qq}

Relevant work on K2Q systems include~\citep{Kotov:2010:TNQ,Zheng:2011:K2Q,Zhao:2011:AGQ,Dror:2013:QQO}.  All these systems follow a template-based approach, and are evaluated in terms of relevance, diversity, and grammatical correctness.  While some differences exist among these systems, all consist of three main steps.  First, they extract question templates from millions of keyword-question pairs by substituting keyword terms in questions with slots, and storing keyword-template pairs in a database $D_{{\langle}q,t{\rangle}}$.  Second, given a new keyword query $q'$, they search similar keyword queries from $D_{{\langle}q,t{\rangle}}$, collect templates related to those similar queries, and instantiate those templates with $q'$ for generating candidate questions.  Finally, a parameterized ranking model is used to calculate the probabilities of those candidate questions being generated by the query $q'$, and to rank all candidate questions.  Instead of template-based methods, we propose to address the K2Q task using state-of-art neural machine translation approaches. 

Other question generation tasks were also addressed in the literature, including converting assertions identified in text (sentences, paragraphs) into question forms,\cite{Agarwal:2011:AQG, Zhou:2017:NQG}. In contrast, our task aims to expand keyword terms into a natural language question.

\subsection{Synthetic Data Generation} 
\label{subsec:sdg}

The idea of automatically generating synthetic data (pseudo test collections) for information retrieval (IR) has attracted some attention in past years~\citep{Azzopardi:2007:BSQ, Berendsen:2013:PTC}.  To the best of our knowledge, utilizing simulated queries for evaluating IR was first suggested by \citet{Azzopardi:2006:ACK}, who also proposed an algorithm for generating simulated queries~\citep{Azzopardi:2006:ACK, Azzopardi:2007:BSQ}.  Their experimental results show that it is possible to generate simulated queries for web search with performance comparable to that of manual queries.
Besides, the idea of generating pseudo test collections was also utilized for the training of supervised (learning-to-rank) retrieval models for web search~\cite{Asadi:2011:PTC} and for ad-hoc search on domain-specific, semi-structured documents~\cite{Berendsen:2012:GPT, Berendsen:2013:PTC}. 
It should be pointed out that automatically generated synthetic training data for deep learning had accomplished a great deal in computer vision~\citep{Handa:2015:SUR, Zhang:2015:SUR, Gan:2015:LDS, Ros:2016:SDL}.  Even though synthetic data is imperfect, these efforts show the feasibility of training robust and effective neural network models with noisy, but very large-scale data.
In this paper, we have proposed a keyword query generation model and developed various filtering mechanisms, in order to create synthetic training data for training neural K2Q models.

\section{Conclusions}
\label{sec:concl}

In this work, we have studied the problem of translating keyword queries to natural language questions using neural approaches.  To the best of our knowledge, this is the first application of neural machine translation methods to the keyword-to-question (K2Q) task.  Perhaps the most innovative aspect of this work is the combination of keyword query generation models combined with various filtering mechanisms to create massive amounts of synthetic data for training neural models.  Our empirical evaluation has demonstrated the effectiveness of our synthetic data generation approach for the K2Q task.  

In this paper, we have generated only a single question for each keyword query, and evaluated it with respect to relevance and grammatical correctness.  The same neural models, however, may also be used to generate a diverse list of questions for a given keyword query, with the help of techniques like beam search~\cite{Freitag:2017:BSS}.  For example, given the keyword query ``\emph{Bible verse about education}," our neural models generated a range of diverse and meaningful questions, including:
\begin{itemize}
	\item ``\emph{What is the fugitive slave verse about education?}"
	\item ``\emph{What is the christ verse about education?}"
	\item ``\emph{What is the sacred verse about education?}"
	\item ``\emph{What does Bible verse say about education?}"
\end{itemize}
In the future, we are interested in generating a diverse set of questions (i.e., the bottom part in Fig.~\ref{fig:ex1}) and comparing these with existing template-based methods with respect to diversity.

In summary, our methods have shown great potential and promise for creating synthetic training data that can be used to train robust neural models; future applications of this idea extend beyond the keyword-to-question task.

\bibliography{arxiv-k2q.bib}
\bibliographystyle{ACM-Reference-Format}

\end{document}